\documentclass{elsart}

\usepackage{epsfig}

\usepackage{amssymb}

\newcommand{\bq}{\begin{equation}}
\newcommand{\eq}{\end{equation}}
\newcommand{\bqa}{\begin{eqnarray}}
\newcommand{\eqa}{\end{eqnarray}}
\newcommand{\ben}{\begin{enumerate}}
\newcommand{\een}{\end{enumerate}}
\newcommand{\bc}{\begin{center}}
\newcommand{\ec}{\end{center}}
\newcommand{\bqb}{\begin{eqnarray*}}
\newcommand{\eqb}{\end{eqnarray*}}

\def\gsim{\gtrsim}
\def\lsim{\lesssim}

\def\ie{{\it i.e. ~}}
\def\eg{{\it e.g. ~}}

\def\pl#1#2#3{ Phys. Lett. ${\bf{#1}}$ (#2) #3}

\def\epj#1#2#3{ Eur. Phys. J. ${\bf{#1}}$ (#2) #3}
\def\np#1#2#3{ Nucl. Phys. ${\bf{#1}}$ (#2) #3}

\begin{document}

\begin{frontmatter}



\title{The processes
$\gamma \gamma \to \gamma \gamma,~ \gamma Z,~ZZ $.}


\author{G.J. Gounaris }

\address{Department of Theoretical Physics, Aristotle
University of Thessaloniki,\\
Gr-54006, Thessaloniki, Greece.\\
E-mail: gounaris@physics.auth.gr}

\author{(Collaboration with
 J. Layssac, P.I. Porfyriadis and F.M. Renard.)}

\begin{abstract}
The Standard Model contributions to the processes
$\gamma \gamma \to \gamma \gamma$, $\gamma Z$ and $ZZ$
at sufficiently high energies,
are found to be helicity conserving and almost purely
imaginary. This is due to the $W$-loop contribution, which
is much bigger than the fermionic ones at such energies.
Thus the structure of
these amplitudes acquires an impressive simplicity at high
energies. Nothing like this appears in other process like \eg~ the
the production of a pair of neutral Higgs bosons.
\end{abstract}

\begin{keyword}
Standard Model \sep  Photon \sep Collider \sep $Z$
\PACS 12.15.Ji \sep 14.70.Bh \sep 14.70.Hp
\end{keyword}
\end{frontmatter}

\vspace*{-15.3cm}
\noindent hep-ph/0008xxx \\
THES-TP 2000/10
\vspace*{14.3cm}

\begin{figure*}
\vspace*{0cm}
\[
\hspace{-1.cm}
\epsfig{file=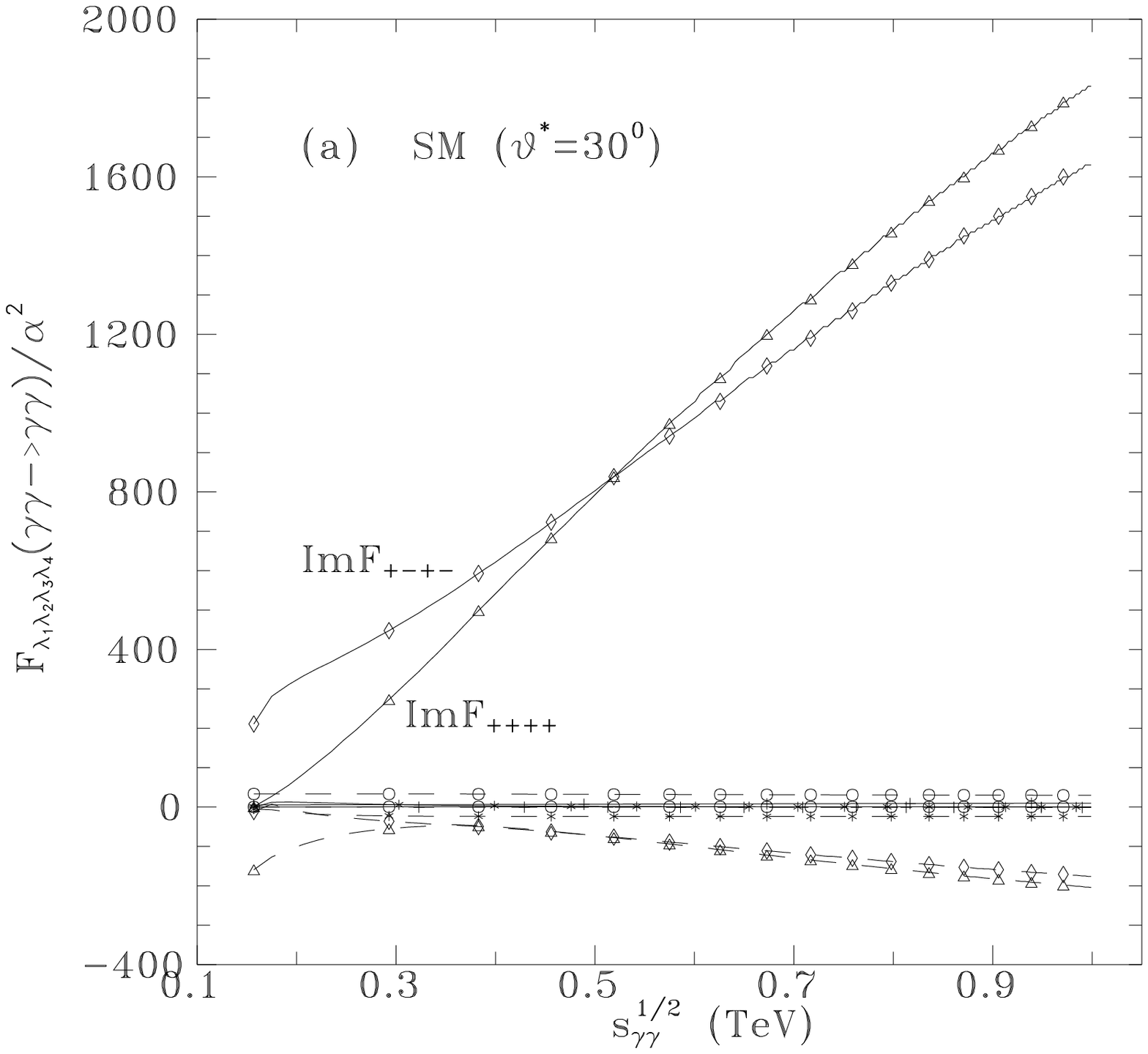,height=6.cm}\hspace{0.5cm}
\epsfig{file=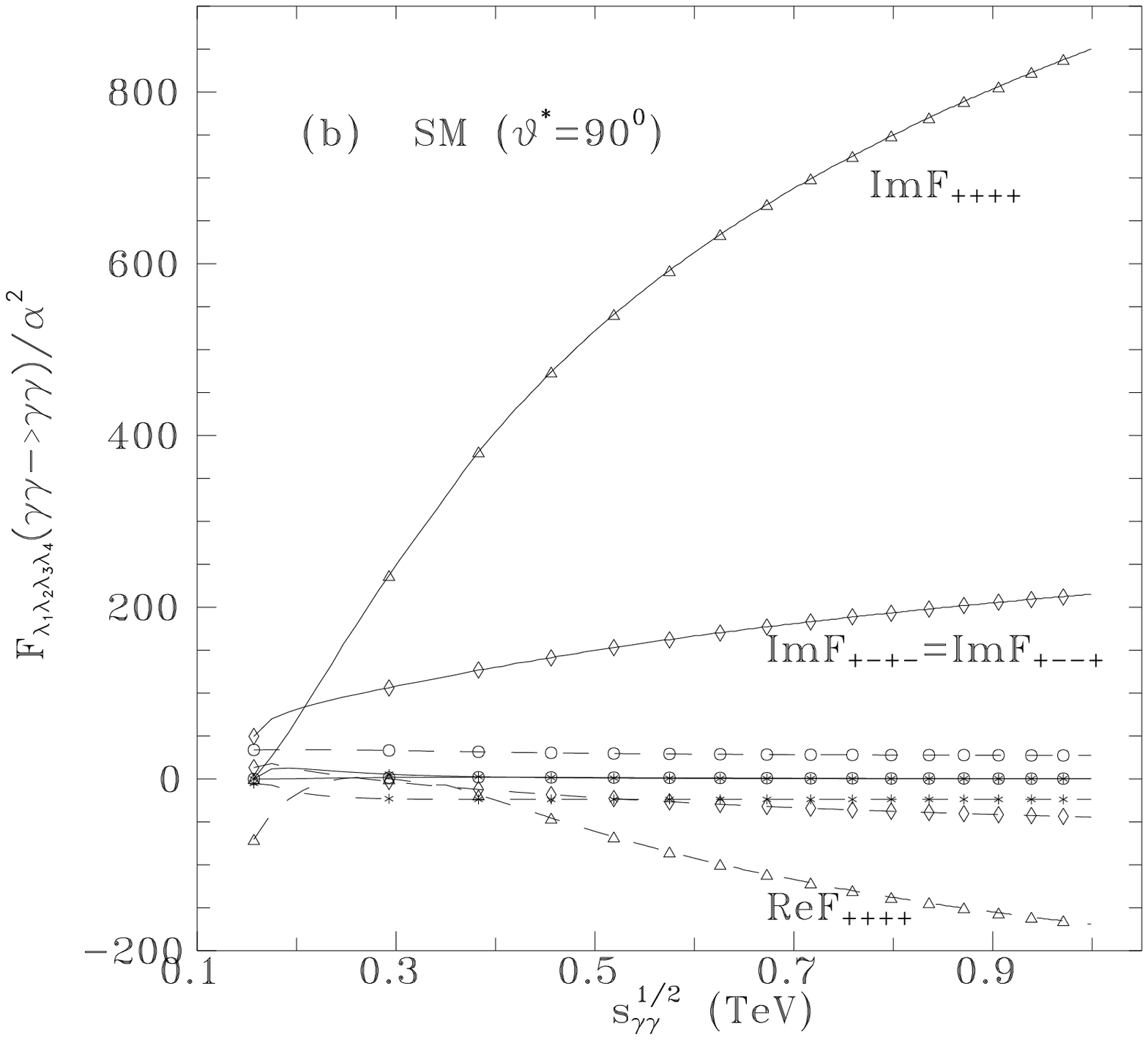,height=6.cm}
\]
\vspace*{-.5cm}
\caption[1]{SM contribution to the $\gamma \gamma \to \gamma \gamma $
helicity amplitudes at $\vartheta^*=30^0$ and $\vartheta^*=90^0$.
Solid (dash) lines describe Imaginary (Real) parts respectively.}
\label{SM-gggg-amp}
\end{figure*}
\begin{figure*}
\vspace*{0.5cm}
\[
\hspace{-1.cm}
\epsfig{file=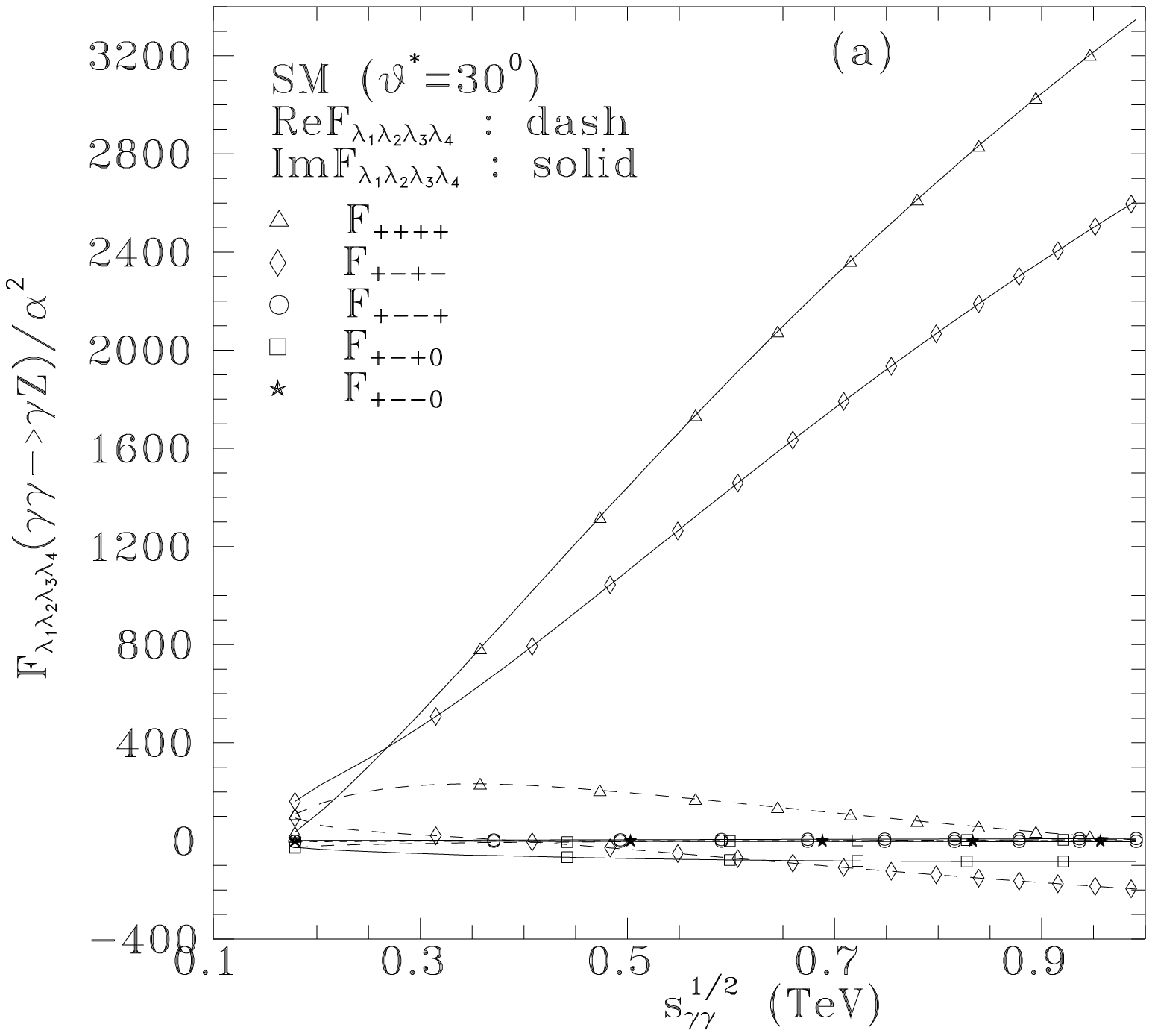,height=6.cm}\hspace{0.5cm}
\epsfig{file=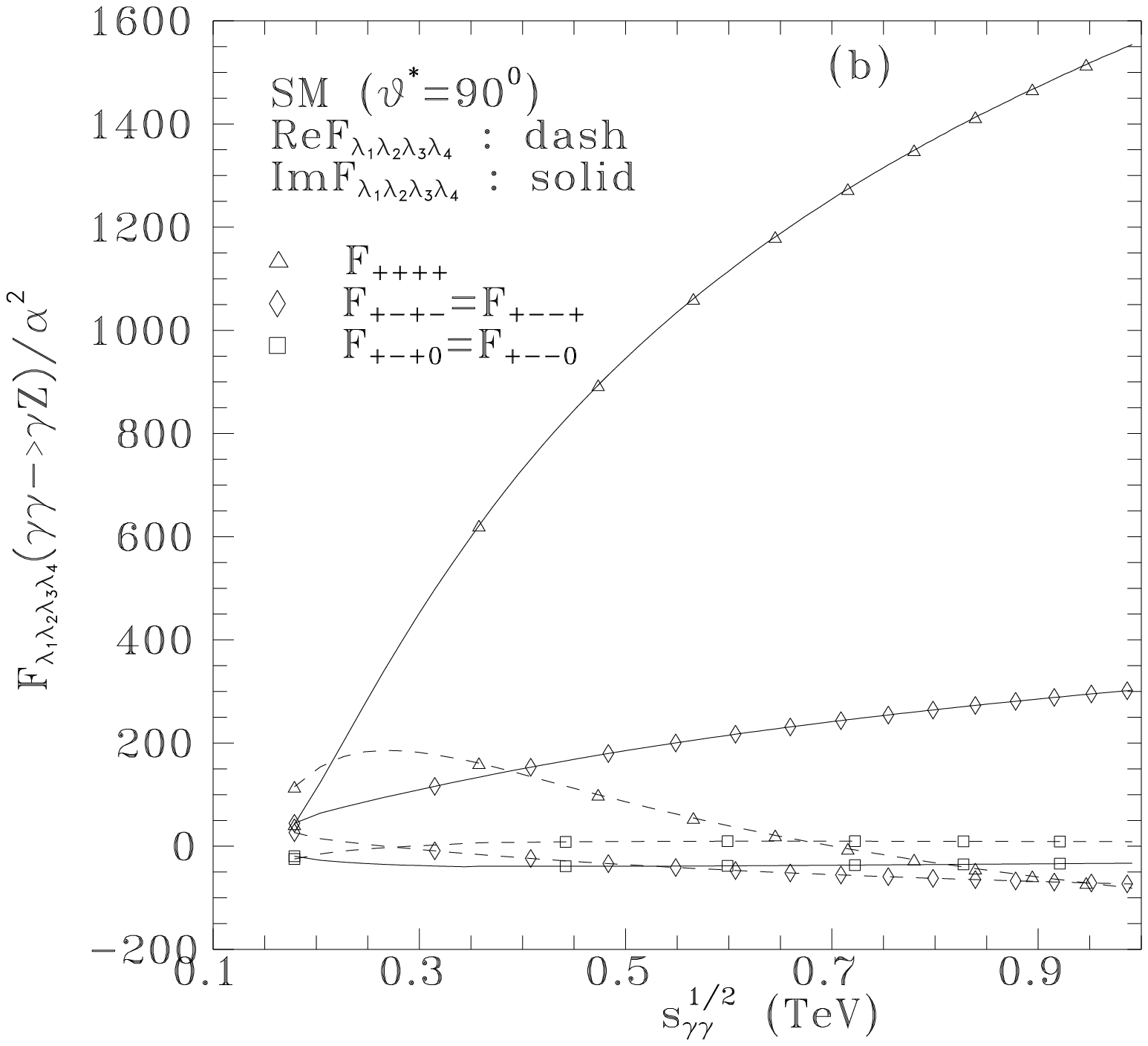,height=6.cm}
\]
\vspace*{-.5cm}
\caption[1]{SM contribution to the $\gamma \gamma \to \gamma Z $
at $\vartheta^*=30^0$ and $\vartheta^*=90^0$.
Solid (dash) lines describe Imaginary (Real) parts respectively.}
\label{SM-gggZ-amp}
\end{figure*}
\begin{figure*}
\vspace*{0.5cm}
\[
\hspace{-1.cm}
\epsfig{file=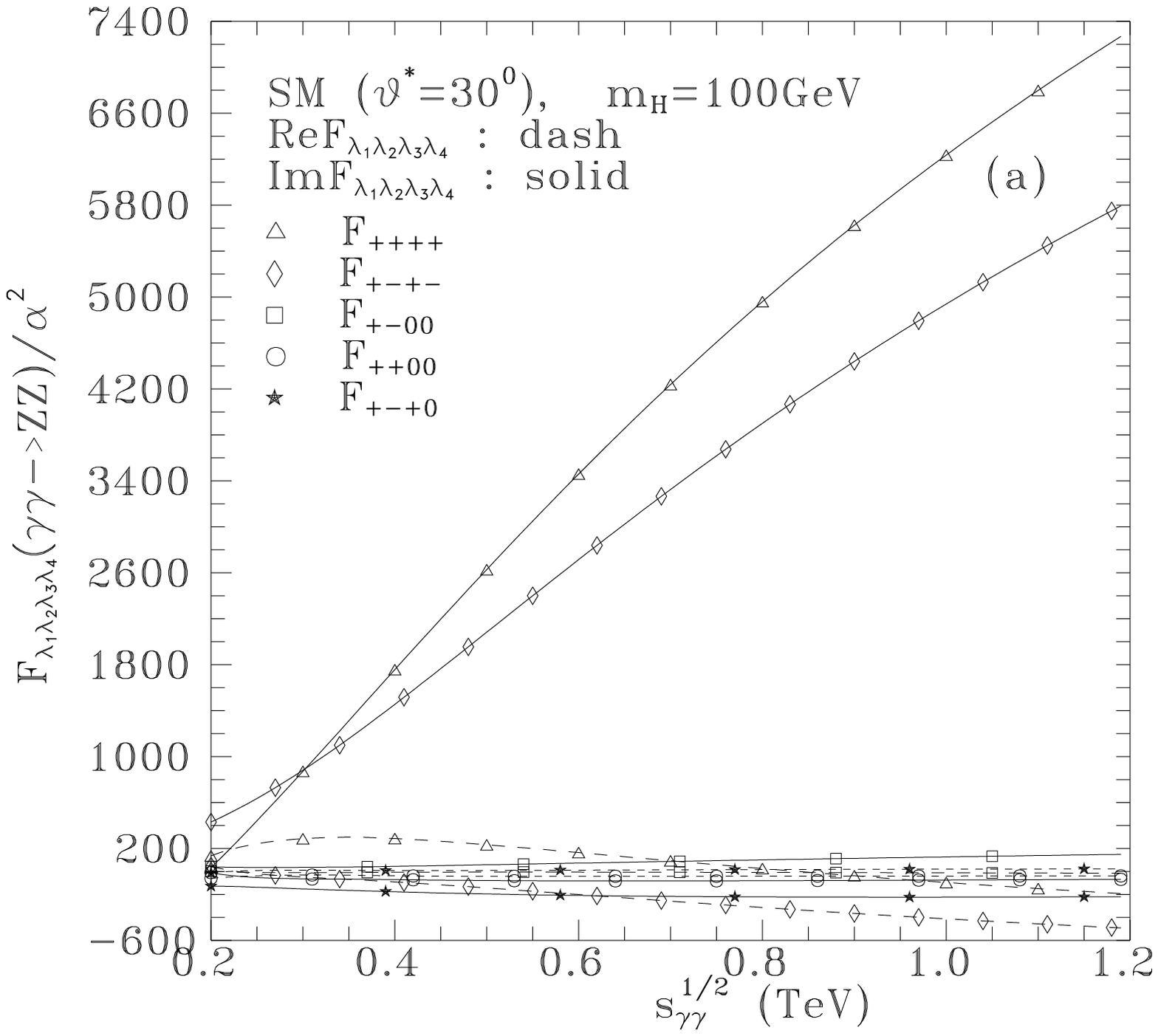,height=6.cm}\hspace{0.5cm}
\epsfig{file=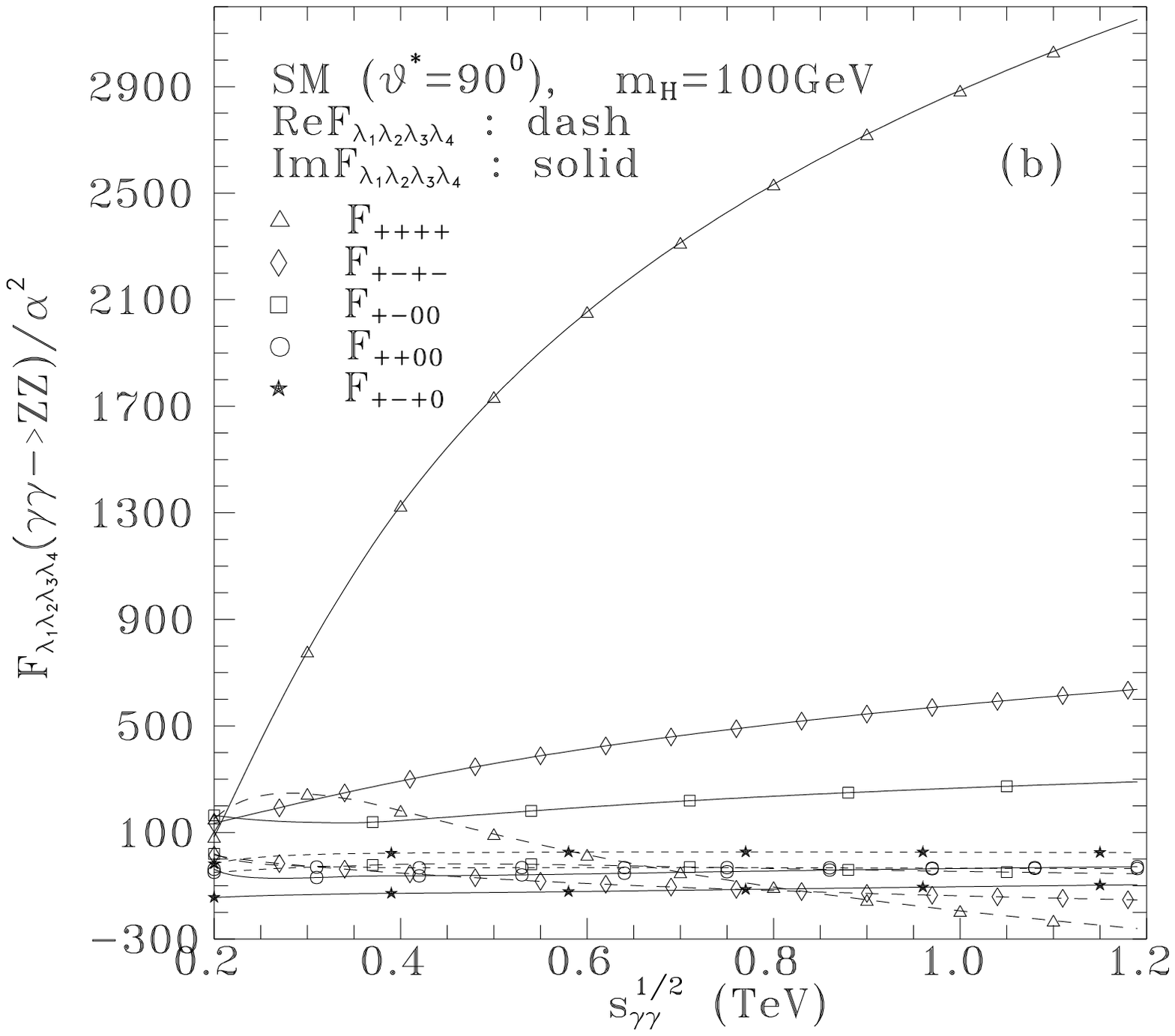,height=6.cm}
\]
\vspace*{-.5cm}
\caption[1]{SM contribution to the $\gamma \gamma \to ZZ $
helicity amplitudes at $\vartheta^*=30^0$ and
$\vartheta^*=90^0$ for $m_H=100GeV$.
Solid (dash) lines describe Imaginary (Real) parts respectively.}
\label{SM-ggZZ-amp}
\end{figure*}

The number of helicity amplitudes that can in principle contribute
to the processes $\gamma \gamma \to \gamma \gamma $, $\gamma Z$,
$ZZ$ is quite large, due to the spin=1 nature of the four
particles involved. In the Standard Model (SM) these processes
receive no tree level contributions, thus getting their
lowest order amplitudes from  1-loop diagrams involving
either quarks and charged leptons, or $W$-bosons. At energies
below \eg $\sim 250 {\rm GeV}$, and for sufficiently large scattering
angles so that the perturbative
1-loop calculation to be reliable,  the various possible helicity
amplitudes for these processes  are more or less on the same footing.
The situation considerably simplifies though
at high energies; \ie at $\sqrt{s} \gsim 250 {\rm GeV}$ for
$(\gamma \gamma \to \gamma \gamma, ~ \gamma Z)$, and at
$\sqrt{s} \gsim 300 {\rm GeV}$ for $ZZ$ production. At such
energies the $W$-loop contribution completely dominates  the
fermionic one; and  only the two helicity conserving amplitudes
$F_{++++}$ and $F_{+-+-}$ remain important\footnote{Together of
course with  those related to it by CP transformations and Bose
statistics.}. Moreover, these predominant amplitudes are almost
purely imaginary. This
can be seen  for $\gamma \gamma$, $\gamma Z$ and $ZZ$ production
 from Figs.\ref{SM-gggg-amp}, \ref{SM-gggZ-amp} and
Fig.\ref{SM-ggZZ-amp} respectively \cite{Jikia}, \cite{gggg},
\cite{gggZ}, \cite{ggZZ}. Please note that the amplitudes not
listed in these figures are at most comparable to the smallest
among the listed ones. \par

The physical reason for the dominant amplitudes becoming almost
purely imaginary at very high energies is not particularly clear.
For some reason, it turns out that the Sudakov-like $log^2$ terms
always cancel at high energies, and only the single-log imaginary
terms remain, and these for the helicity conserving amplitudes only.
We have also looked at the  SM contribution to the $\gamma \gamma
\to h^0 h^0$, where $h^0$ is the SM Higgs particle, as well to
$\gamma \gamma \to A^0 A^0$ ($A^0$ being the CP odd Higgs particle
in SUSY models), and no particular dominance  for the imaginary parts
of any amplitudes was observed \cite{ggAA}.

Concerning in particular the results in Fig.\ref{SM-ggZZ-amp} for
$ZZ$ production, we should remark that they only apply for the light
Higgs case. In that figure $m_h \sim 100\rm GeV$ is used, which
should,  be responsible for the tiny $F_{++00}$ amplitude. For
Higgs masses at the TeV scale, the importance of $F_{++00}$  amplitude
should obviously increase.\par

\begin{figure*}
\vspace*{0.5cm}
\[
\epsfig{file=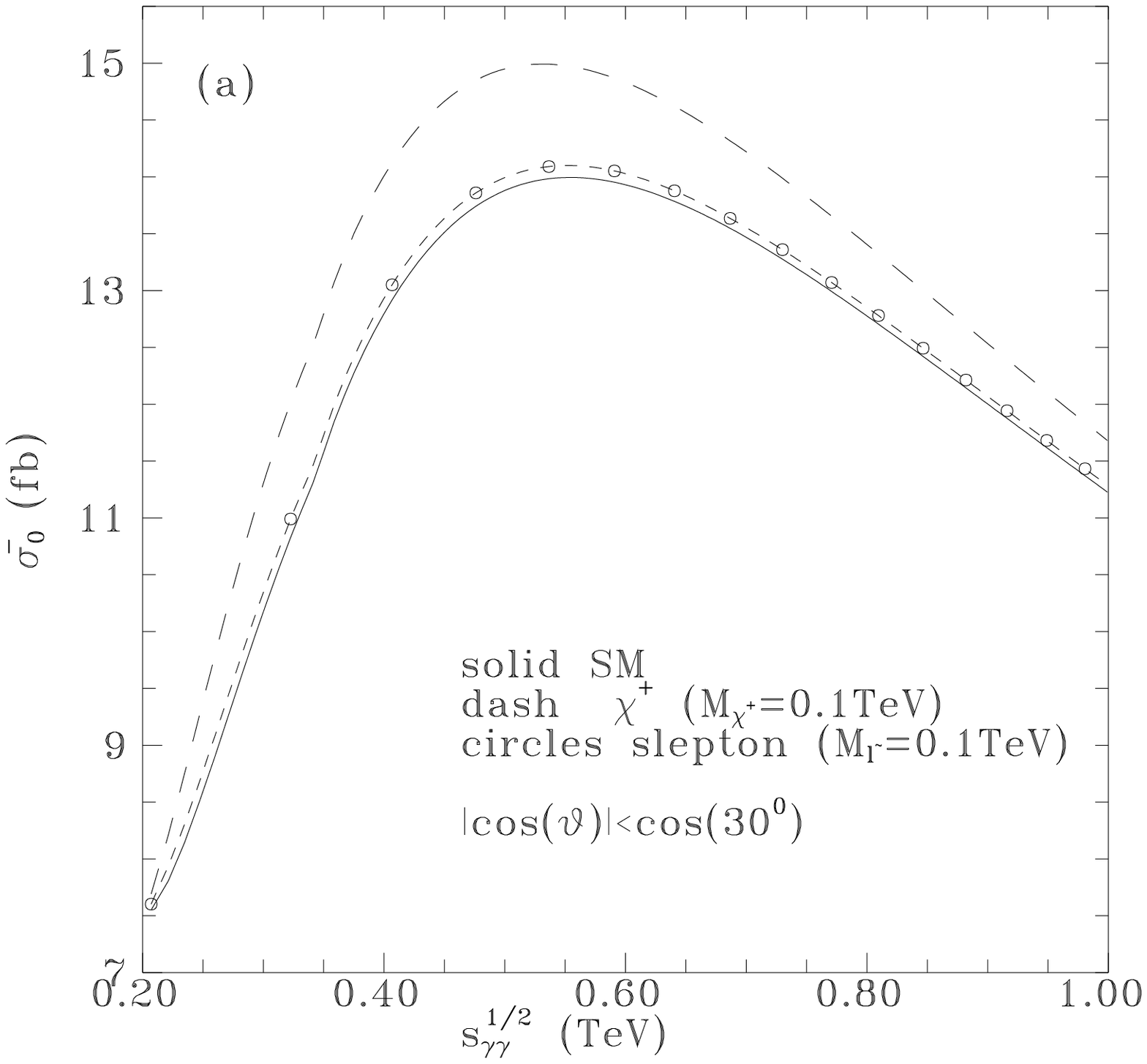,height=7.cm}
\]
\vspace*{-.5cm}
\caption[1]{SM (solid)  and the effect of including
various SUSY (dash)
contributions to the unpolarized
$\gamma \gamma \to \gamma \gamma  $ cross section integrated
for center of mass angles $30^0 \leq \vartheta^* \leq 90^0$.}
\label{sig0-gggg}
\end{figure*}

\begin{figure*}
\vspace*{0.5cm}
\[
\epsfig{file=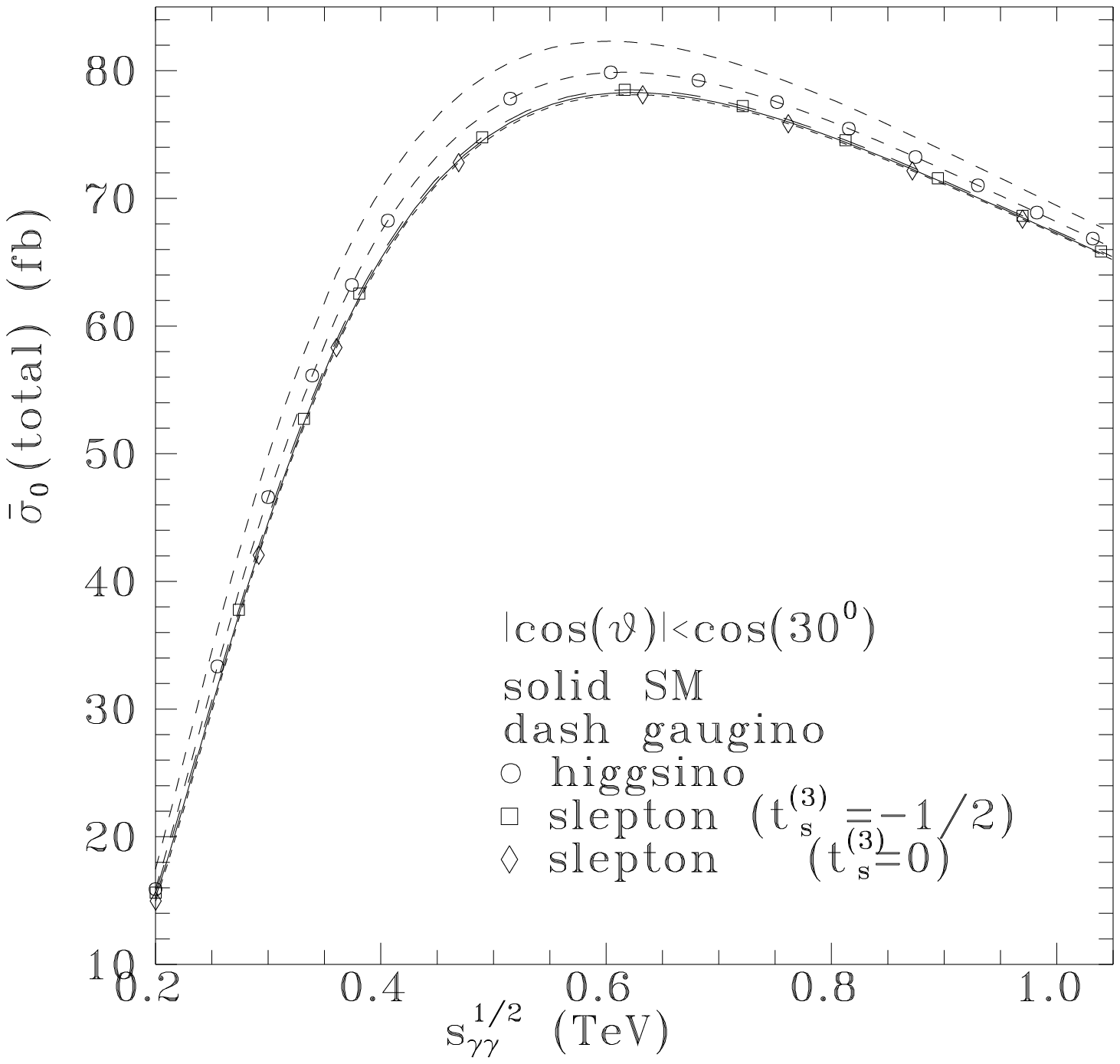,height=7.cm}
\]
\vspace*{-.5cm}
\caption[1]{SM (solid)  and the effect of including various SUSY (dash)
 contributions to the unpolarized
$\gamma \gamma \to \gamma Z  $ cross section integrated
for center of mass angles $30^0 \leq \vartheta^* \leq 90^0$.}
\label{sig0-gggZ}
\end{figure*}

\begin{figure*}
\vspace*{0.5cm}
\[
\epsfig{file=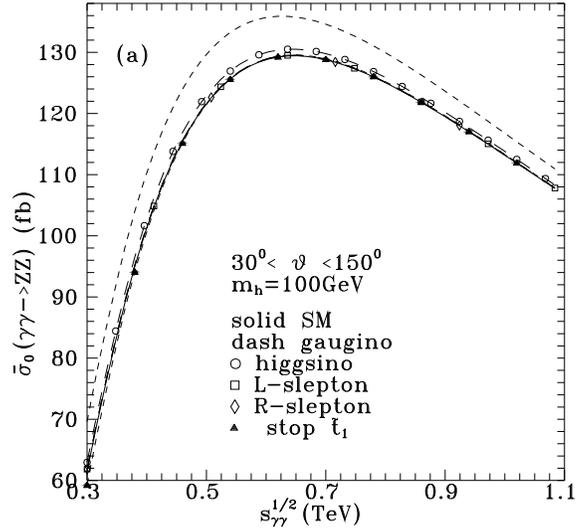,height=7.cm}
\]
\vspace*{-.5cm}
\caption[1]{SM (solid)  and effect of including various SUSY (dash)
contributions to the unpolarized
$\gamma \gamma \to ZZ  $ cross section integrated
for center of mass angles $30^0 \leq \vartheta^* \leq 90^0$.
The parameters entering the Higgs pole contribution are chosen
in the decoupling SUSY regime.}
\label{sig0-ggZZ}
\end{figure*}

Thus, at high energies and for a small mass of the lightest Higgs,
the processes ($\gamma \gamma \to \gamma \gamma $, $\gamma Z$,
$ZZ$) acquire a striking simplicity, at high energies.
The corresponding un-polarized cross sections, integrated in the range
$30^0 \leq \vartheta^* \leq 150^0$ are shown by the solid lines in
Figs.\ref{sig0-gggg}, \ref{sig0-gggZ}, \ref{sig0-ggZZ}. Notice
that in deriving these results $\alpha=1/128$ has been used. If
$\alpha =1/137$ is thought more appropriate, then the results for the
cross sections should be multiplied by $\sim 0.76$. In any case,
such cross sections for unpolarized as well as for polarized
beams\footnote{See  \cite{gggg}, \cite{gggZ}, \cite{ggZZ}.}
should be measurable  if a $\gamma \gamma $ Collider
($ LC_{\gamma \gamma}$) is ever
built with the anticipated Luminosity \cite{Ginzburg},
\cite{Telnov}.

The aforementioned effect should  be useful in searching for
New Physics (NP) in an $ LC_{\gamma \gamma}$. Please
notice, that in order to get appreciable interference between the NP
effect to ($\gamma \gamma \to \gamma \gamma ,~ \gamma Z, ~ ZZ$)
and the predominantly imaginary and helicity conserving SM
amplitudes, we always need to be above the threshold for their
direct production. Thus in \cite{gggg}, \cite{gggZ},
\cite{ggZZ} we have explored the possibility to use the above
processes in order to get additional independent
 information that should help
identifying  the nature of possible SUSY candidates,
 that may also be directly
produced. In this respect, it should be remarked
that  the virtual effects induced by
the various SUSY particles depend on  different sets of
parameters than those affecting their decay. We have found that
the experimental study of $\gamma \gamma \to \gamma \gamma ,` \gamma Z, ~ ZZ$
may be particularly useful  for chargino-type candidates, provided
their mass is $\lsim 200 GeV$. This can also be guessed
from Figs.\ref{sig0-gggg}, \ref{sig0-gggZ},
\ref{sig0-ggZZ} in which the chargino effect is always the biggest
among those induced by possible SUSY candidates.

Another use, particularly of the process
$\gamma \gamma \to \gamma \gamma $, which has repeatedly appeared in
the recent literature,  is in order to look  for effects
due to strings of gravitons
exchanged between the photon pairs, in case extra  large dimensions
might exist. Unfortunately, since this NP contribution is mainly
real, there is no appreciable NP-SM interference, and the NP effect is
mainly  sensitive to the square of the NP amplitude. In spite of
this, the quoted sensitivity appears appreciable \cite{Choudhury}


\clearpage


\begin{thebibliography}{00}


%
\bibitem{Jikia} G. Jikia and A. Tkabladze \pl{B323}{1994}{433};
ibid \pl{B332}{1994}{441}; G. Jikia \np{B405}{1993}{24}.
%
\bibitem{gggg} G.J. Gounaris, P.I. Porfyriadis, F.M.
Renard, hep-ph/9812378, \pl{B452}{1999}{76},
\pl{B464}{1999}{350} (E); G.J. Gounaris, P.I. Porfyriadis, F.M.
Renard,  hep-ph/9902230, \epj{C9}{1999}{673}.
%
\bibitem{gggZ}G.J. Gounaris, J. Layssac, P.I. Porfyriadis and
 F.M. Renard,  hep-ph/9904450, \epj{C10}{1999}{499}.
%
\bibitem{ggZZ}G.J. Gounaris, J. Layssac, P.I. Porfyriadis and
 F.M. Renard,  hep-ph/9909243, \epj{C13}{2000}{79}.
%
\bibitem{ggAA} G.J. Gounaris and P.I. Porfyriadis,
hep-ph/0007110.
%
\bibitem{Ginzburg}
I.~F.~Ginzburg, G.~L.~Kotkin, S.~L.~Panfil, V.~G.~Serbo and V.~I.~Telnov,
Nucl.\ Instrum.\ Meth.\  {\bf 219} (1984) 5;
V.~I.~Telnov,
Nucl.\ Instrum.\ Meth.\  {\bf A294} (1990) 72.
%
\bibitem{Telnov} V. Telnov, these proceeding,
http://www.desy.de/ gg2000,  June 14-17, 2000,
DESY Hamburg, Germany, to appear in Nucl.Instr. \& Meth. A.
%
\bibitem{Choudhury} see \eg S.R. Choudhury, A. Cornell and G.C.
Joshi, hep-ph/0007347.

\end{thebibliography}
\end{document}